\UseRawInputEncoding 
\documentclass[superscriptaddress,aps,prl,reprint,twocolumn,amsmath,amssymb,showpacs,]{revtex4-2}
\usepackage{graphicx}  
\usepackage{mathptmx}
\usepackage{epstopdf}
\usepackage{dcolumn}   
\usepackage{bm}        
\usepackage{amssymb}   
\usepackage{amsmath}   
\usepackage{amsthm}
\usepackage{chemarrow}
\usepackage{color}
\usepackage{mathrsfs}
\usepackage{float}
\usepackage{physics}
\usepackage{bbold}
\usepackage{bbm}
\usepackage[a4paper,colorlinks=true,
linkcolor=blue,citecolor=blue,
pdfauthor={ },
pdftitle={ },
pdfsubject={ },
pdfkeywords={ }]{hyperref}

\theoremstyle{plain}
\newtheorem*{theorem*}{Theorem}

\begin{document}


\title{Enhanced quantum sensing with amplification and deamplification}

\date{\today}

\author{Min Jiang}
\email[]{These authors contributed equally to this work}
\affiliation{
CAS Key Laboratory of Microscale Magnetic Resonance and School of Physical Sciences, University of Science and Technology of China, Hefei, Anhui 230026, China}
\affiliation{
CAS Center for Excellence in Quantum Information and Quantum Physics, University of Science and Technology of China, Hefei, Anhui 230026, China}

\author{Yushu Qin}
\email[]{These authors contributed equally to this work}
\affiliation{
CAS Key Laboratory of Microscale Magnetic Resonance and School of Physical Sciences, University of Science and Technology of China, Hefei, Anhui 230026, China}
\affiliation{
CAS Center for Excellence in Quantum Information and Quantum Physics, University of Science and Technology of China, Hefei, Anhui 230026, China}

\author{Yuanhong Wang}
\affiliation{
CAS Key Laboratory of Microscale Magnetic Resonance and School of Physical Sciences, University of Science and Technology of China, Hefei, Anhui 230026, China}
\affiliation{
CAS Center for Excellence in Quantum Information and Quantum Physics, University of Science and Technology of China, Hefei, Anhui 230026, China}

\author{Ying Huang}
\affiliation{
CAS Key Laboratory of Microscale Magnetic Resonance and School of Physical Sciences, University of Science and Technology of China, Hefei, Anhui 230026, China}
\affiliation{
CAS Center for Excellence in Quantum Information and Quantum Physics, University of Science and Technology of China, Hefei, Anhui 230026, China}

\author{Xinhua Peng}
\email[]{xhpeng@ustc.edu.cn}
\affiliation{
CAS Key Laboratory of Microscale Magnetic Resonance and School of Physical Sciences, University of Science and Technology of China, Hefei, Anhui 230026, China}
\affiliation{
CAS Center for Excellence in Quantum Information and Quantum Physics, University of Science and Technology of China, Hefei, Anhui 230026, China}
\affiliation{
Hefei National Laboratory, University of Science and Technology of China,
Hefei 230088, China}

\author{Dmitry Budker}
\affiliation{Helmholtz-Institut, GSI Helmholtzzentrum f{\"u}r Schwerionenforschung, Mainz 55128, Germany}
\affiliation{Johannes Gutenberg University, Mainz 55128, Germany}
\affiliation{Department of Physics, University of California, Berkeley, CA 94720-7300, USA}

\begin{abstract}
Quantum sensing is a fundamental building block of modern technology that employs quantum resources and creates new opportunities for precision measurements.
However, previous methods usually have a common assumption that detection noise levels should be below the intrinsic sensitivity provided by quantum resources.
Here we report the first demonstration of Fano resonance between coupled alkali-metal and noble gases through rapid spin-exchange collisions.
The Fano resonance gives rise to two intriguing phenomena: spin amplification and deamplification,
which serve as crucial resources for enhanced sensing.
Further we develop a novel scheme of quantum sensing enhanced by amplification and deamplification,
with relaxed requirements on the detection noise.
The coupled systems of alkali-metal and noble gases act as amplifiers or de-amplifiers, enabling to extract small signals above the detection noise before final detection.
We demonstrate magnetic-field measurement about 54 decibels below the photon-shot noise,
which outperforms the state-of-the-art squeezed-light technology and realizes femtotesla-level sensitivity.
Our work opens new avenues to applications in searches for ultralight dark matter with sensitivity well beyond the supernova-observation constraints.
\end{abstract}

\maketitle

\begin{figure*}[t]  
	\makeatletter
	\def\@captype{figure}
	\makeatother
	\begin{center}
	    \includegraphics[scale=1.075]{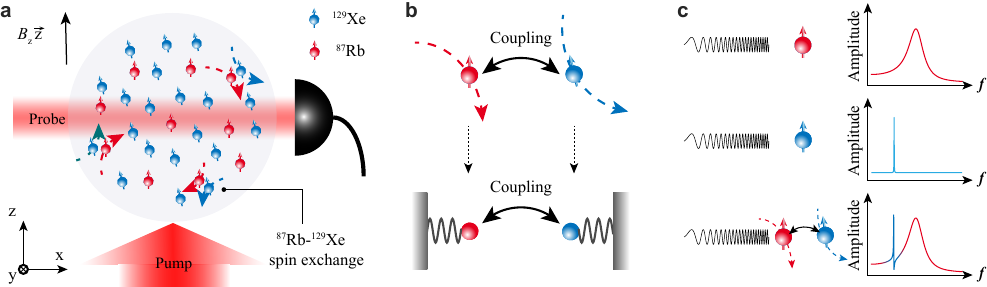}
	\end{center}
	\caption{\textbf{Principle of enhanced sensing using Fano resonance.} \textbf{a}, Experimental setup. The key element is a 0.5\,cm$^3$ vapor cell containing 20\,torr
$\rm ^{129}Xe$, 250\,torr $\rm N_2$, and a droplet of isotopically enriched $\rm^{87}Rb$. 
$\rm ^{129}Xe$ spins are polarized and coupled with $\rm^{87}Rb$ by Fermi-contact collisions. $\rm^{87}Rb$ atoms are polarized with a circularly-polarized pump laser light tuned to the D1 line and probed with a linearly-polarized probe laser light detuned to higher frequencies by 110\,GHz from the D2 line.
\textbf{b}, Coupled oscillator model. The coupled alkali-metal-noble-gas spins can be seen as two coupled harmonic oscillators.
\textbf{c}, Schematic of Fano resonance. $\rm^{87}Rb$ provides a ``continuum'' response profile with a broad resonance linewidth ($\Gamma_a \approx 30$\,kHz) and $\rm^{129}Xe$ provides a ``discrete-system'' resonance with narrow linewidth ($\Gamma_b \approx 7$\,mHz).
The response profile for individual $\rm^{87}Rb$ or $^{129}$Xe is symmetric (top and middle).
Interference between their response signals leads to the asymmetric Fano profile in the spectral vicinity of the $\rm ^{129}Xe$ resonance frequency (bottom).}
	\label{Figure1}
\end{figure*}

Quantum sensing uses quantum resources to enable or enhance sensing,
and has created new opportunities for a broad range of metrological tasks\,\cite{degen2017quantum,giovannetti2011advances,clerk2010introduction}.
Examples include numerous quantum technologies, ranging from electromagnetic field sensing\,\cite{kominis2003subfemtotesla,budker2007optical,bao2020spin,jing2020atomic},
frequency standards\,\cite{gill2005optical}, and
gravity-wave detection\,\cite{schnabel2010quantum} to searches for new physics beyond the standard model\,\cite{afach2021search,jiang2021search,safronova2018search,backes2021quantum}.
However, previous works usually required detection noise added during precision measurements below the limits set by the intrinsic sensitivity provided by these quantum resources.
In fact, numerous susceptibilities to detection noise,
such as environmental noise, sensor losses, and shot noise, tend to hamper quantum advantage and prevent the realization of practical quantum-enhanced sensors.
In addition, there are numerous applications of quantum sensors in challenging environments\,\cite{fu2020sensitive}, for example, in cities with high electromagnetic noise or in space with high radiation level,
where low-noise detection is difficult.
To address this, 
there has been dedicated work on enhancing detection sensitivity with cavity-aided measurements\,\cite{hosten2016quantum,wu2022enhanced},
quantum lock-in detection\,\cite{boss2017quantum},
phase-transition enhancement\,\cite{ding2022enhanced}, and filtering techniques\,\cite{kong2020measurement,abbott2004analysis}.

Here, we describe the concept and the realization of enhanced quantum sensing with amplification and deamplification that relaxes the stringent low-noise detection requirements.
We observe a previously unexplored Fano resonance in alkali metal and noble gas coupled by rapid spin-exchange collisions.
The Fano resonance is the key ingredient of our enhanced metrology due to the emerging phenomena of both resonant amplification and deamplification. 
As a first application,
we demonstrate enhanced magnetic-field sensing assisted with amplification of the input signal that is 54\,dB below the photon-shot noise and with a suppression (deamplification) of the magnetic background noise by about 24\,dB.
In contrast to the previous work that exploited squeezed light,
our method that does not involve squeezing is easier to implement, while it achieves substantial sensitivity improvement for quantum sensing.
In addition, our findings unveil a new self-compensating mechanism so that magnetic-field noise does not contribute to the final signal.
In particular, the present mechanism is perfectly suited to understanding previously-established self-compensated comagnetometry\,\cite{kornack2002dynamics,vasilakis2009limits,liu2022dynamics} operating in the near-zero frequency,
whereas our work can operate in a broader frequency range for self-compensating.
We further discuss prospects for quantum-enhanced sensing using coupled atomic gases and its potential utility in applications, for example, non-Hermitian metrology\,\cite{kononchuk2022exceptional,chu2020quantum,cao2020reservoir} and searches for new physics\,\cite{afach2021search,jiang2021search,safronova2018search,backes2021quantum}.

Our quantum sensing experiments are carried out in a warm vapor cell that contains $^{87}$Rb and $^{129}$Xe gases. These two atomic gases interact with each other through spin-exchange collisions that are divided into incoherent and coherent processes\,\cite{katz2022quantum,shaham2022strong}.
Figure\,\ref{Figure1}(a) schematically illustrates the main steps of the measurement.
The $\rm^{129}Xe$ spins are polarized along $z$ via incoherent collisions with optically-polarized $\rm^{87}Rb$ atoms\,\cite{walker1997spin}.
The coherent process leads to mutual precession between the two spins,
where alkali-metal spins experience an additional effective magnetic field $\lambda \textbf{M}^b (t)$ and noble-gas nuclear spins experience $ \lambda \textbf{M}^a (t)$ as well.
Here $\lambda = 8 \pi \kappa_{0}/3$ and $\kappa_0 \approx 540$ is the Fermi-contact enhancement factor for $\rm^{87}Rb$ and $\rm^{129}Xe$\,\cite{walker1997spin}.
To make the coherent dynamics of the hybrid system more intuitive,
we use Holstein-Primakoff transformation, i.e., $\hat{a}=(M^a_x+iM^a_y)/\sqrt{2\gamma_a M^a_z}$ and $\hat{b}=(M^b_x+iM^b_y)/\sqrt{2\gamma_b M^b_z}$\,\cite{ressayre1975holstein}.
The measured field is assumed to be so weak that the excited transverse field $M^a_x$ and $M^a_y$ satisfies $M^a_x,M^a_x\ll M^a_z$.
The dynamics of spin excitations $\hat{a}, \hat{b}$ can be modeled by (Supplementary Section\,\uppercase\expandafter{\romannumeral1})
\begin{equation}
    \partial_t\begin{pmatrix}
        \hat{a}\\
        \hat{b}
    \end{pmatrix}=
    i\begin{pmatrix}
        \omega_a+i\Gamma_a & -J\\
        -J & \omega_b+i\Gamma_b
    \end{pmatrix}
    \begin{pmatrix}
         \hat{a}\\
        \hat{b}
    \end{pmatrix}+
    \begin{pmatrix}
         h_a\\
         h_b
    \end{pmatrix},
    \label{eq1}
\end{equation}
where $\omega_a=\gamma_a(B_z+\lambda M^b_z)$ and $\omega_b=\gamma_b(B_z+\lambda M^a_z)$ denotes the Larmor frequency of the alkali-metal spin and noble-gas spin, respectively.
Here $\gamma_{a,b}$ is gyromagnetic ratio and $\Gamma_{a,b}$ is decoherence rate.
The bidirectional coupling between $\hat{a}$ and $\hat{b}$ excitations is described with $J=\lambda\sqrt{\gamma_a\gamma_b M^a_z M ^b_z}$ (in our experiment, $J\approx 30$ Hz),
which emerges from the mutual precession between $^{87}$Rb and $^{129}$Xe gases.
The last term with $h_{a,b}$ represents the excitations induced by the measured field (Supplementary Section\,\uppercase\expandafter{\romannumeral1},\uppercase\expandafter{\romannumeral6}), such as ordinary magnetic field and pseudo-magnetic field from inertial rotation or new physics.
The $\rm^{129}Xe$ spin excitation is read out by optical probing the real part of $\rm^{87}Rb$ excitation\,\cite{walker1997spin}.

At the core of our method lies the Fano resonance in coupled alkali-metal-noble-gas spins. The coupled spin excitations described in Eq.\,(\ref{eq1}) can be seen as two coupled oscillators that can exchange energy with each other via bidirectional Fermi-contact collisions [Fig.\,\ref{Figure1}(b)].
In general, Fano resonance is an intriguing phenomenon in coupled oscillators that has been widely investigated in photonics and nuclear scattering\,\cite{limonov2017fano,miroshnichenko2010fano,luk2010fano}. To appear as Fano resonance, a discrete-level system and a continuum-level system should exist\,\cite{miroshnichenko2010fano}. Accordingly, $\rm^{129}Xe$ nuclear spins have a sharp magnetic resonance line ($\Gamma_b \approx 7$\,mHz) and can be seen as a discrete system, whereas $\rm^{87}Rb$ spins have a broad resonance line ($\Gamma_a \approx 30$\,kHz) and can be approximated as a continuum system. As a consequence of their distinct resonance lines, the $\rm^{129}Xe$ response phase exhibits a $\pi$ jump at its resonance and, by contrast, the $\rm^{87}Rb$ response phase varies slowly.
The interference between $^{87}$Rb and $^{129}$Xe spin excitations induced by the measured field results in the asymmetric $\rm^{87}Rb$ response profile [Fig.\,\ref{Figure1}©] in the spectral vicinity of $\rm^{129}Xe$ resonance frequency.
We show that the $\rm^{87}Rb$ power spectral response can be well described as a Fano profile (Supplementary Section\,\uppercase\expandafter{\romannumeral3})
\begin{equation}  F(\epsilon)=\mathscr{A}(\epsilon)\dfrac{(q+\epsilon)^2}{1+\epsilon^2}+\mathscr{B}(\epsilon),
    \label{eq2}
\end{equation}
where $\epsilon=(\omega-\widetilde{\omega}_{b})/{\widetilde{\Gamma}_b}=2\pi(\nu-\widetilde{\nu}_{b})/{\widetilde{\Gamma}_b}$, $\widetilde{\omega}_{b}$ and
 $\widetilde{\Gamma}_b$ is the dressed $\rm^{129}Xe$ Larmor frequency and decoherence rate via the bidirectional $\rm^{87}Rb$-$\rm^{129}Xe$ coupling (see below).
In the first term, $q$ denotes the Fano parameter, which affects the symmetry of the profile.
The first term in Eq.\,(\ref{eq2}) is asymmetric when the Fano parameter $q$ is non-zero,
where the profile shape reverses when the sign of $q$ changes.
When $q$ is close to zero, the profile becomes symmetric.
$\mathscr{A}(\epsilon)$ and $\mathscr{B}(\epsilon)$ depending on the coupled gases are symmetric and their details are provided in Supplementary Section\,\uppercase\expandafter{\romannumeral3}.

\begin{figure}[t]  
	\makeatletter
 \centering
	\def\@captype{figure}
	\makeatother
	\includegraphics[scale=0.805]{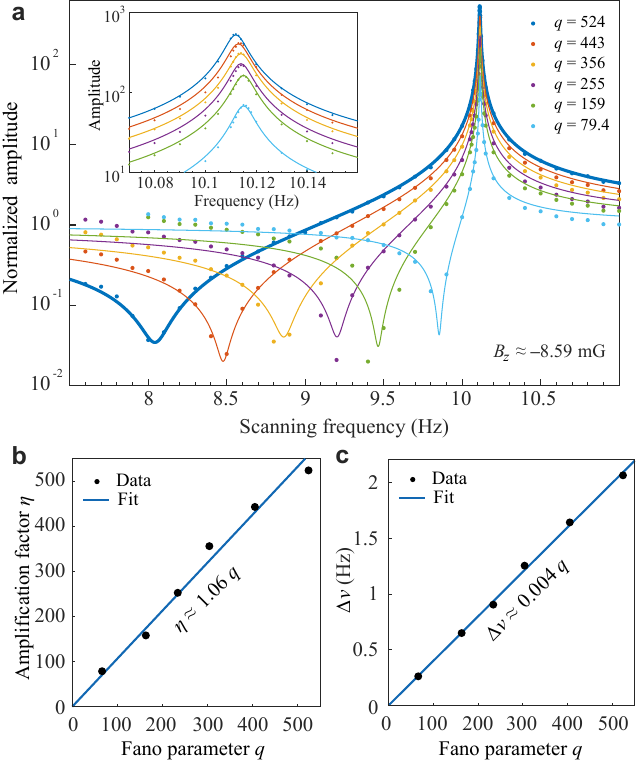}
	\caption{\textbf{Demonstration of Fano resonance, amplification, and deamplification.} \textbf{a}, Fano response profile as a function of the frequency of $y$-directed measured field. $B_z$ is set to about -8.59\,mG as an example. The data are well fit with theoretical Fano profile with Fano parameter $q$, which is modified by applying a periodic magnetic field (Supplementary Section\,\uppercase\expandafter{\romannumeral8}). For each Fano profile, there exists an amplification regime with large response above one and a deamplification regime with response below one. \textbf{b} and \textbf{c}, Demonstration of amplification and deamplification with respect to Fano parameter. The data show that the amplification $\eta$ at $\rm^{129}Xe$ Larmor frequency is nearly equal to the Fano parameter $q$ and the deamplificaton point is linearly dependent on the Fano parameter with the slopes are 1.06\,$\pm$\,0.08 and 0.04\,$\pm$\,0.0001 (95\% confidence interval), respectively (see text). }
	\label{Figure2}
\end{figure}

As a first illustration of Fano resonance,
the external magnetic field is set as $B_z\approx -8.59$\,mG,
corresponding to ${\nu}_{b}\approx 10.11$\,Hz.
We apply an oscillating test field along $y$ and
scan its oscillation frequency around ${\nu}_{b}$, and
record the frequency-response signal. 
The measurement duration of each point in our experiment is 180\,s.
Figure\,\ref{Figure2}(a) shows the experimental data.
For example,
the dark-blue profile is indeed asymmetric and well fit with a Fano profile with Fano parameter $q\approx524$.
In order to observe the change of the Fano profile with the Fano parameter $q$,
we change the Fano parameters by applying a periodic magnetic field on the $^{87}$Rb-$^{129}$Xe vapor cell\,\cite{jiang2022floquet,jiang2021floquet}, 
where the periodic magnetic field is along $z$ direction and has a frequency of 3\,Hz.
The corresponding profiles also show good agreement with the predicted Fano profile.
As $q$ decreases, the maximum value becomes smaller and the frequency of the minimum value moves toward the resonance frequency.

We observe magnetic-field amplification and deamplification,
which are critical resources to bypass the low-noise detection requirements for quantum sensing.
There are two distinct parts of the Fano profile. In one part,
there is constructive interference corresponding to signal enhancement and in the other, there is destructive interference corresponding to signal suppression. These are referred to as ``amplification" and ``deamplification" regimes, respectively.
When the measured frequency $\nu$ is near $\widetilde{\nu}_{b}$ (corresponding to $\epsilon\approx 0$),
the response amplitude in the Fano profile is greatly enhanced and reaches a maximum.
In this case, the effective field $\lambda \mathbf{M}^b$ produced by $^{129}$Xe spins is significantly larger than the measured field,
with an amplification factor of $\eta=\gamma_b  \lambda M^b_z/(2\widetilde{\Gamma}_b$) (Supplementary Section\,\uppercase\expandafter{\romannumeral4}).
For example, the dark-blue response profile in Fig.\,\ref{Figure2}(a) gives $\eta\approx524$,
where the measured field is effectively pre-amplified by at least two orders of magnitude before detection.
As a result, the present amplification offers the capability of extracting small signal below the detection noise level.
We further demonstrate (Supplementary Section\,\uppercase\expandafter{\romannumeral4}) that the amplification factor $\eta$ is equal to the Fano parameter $q$. This is confirmed
by the measurement 
[Fig.\,\ref{Figure2}(b)].
Importantly, this reveals a connection between Fano resonance and amplification that have previously been considered as two distinct phenomena.
On the other hand,
the deamplification with a minimum response occurs at $\epsilon\approx-q$,
according to Eq.\,(\ref{eq2}).
\mbox{Figure}\,\ref{Figure2}(c) shows that the deamplification frequency shifts from the $^{129}$Xe resonance is \mbox{$\Delta v\approx (2\pi)^{-1} q \widetilde{\Gamma}_b$} (Supplementary Section\,\uppercase\expandafter{\romannumeral3}).
The fit to our data gives $\widetilde{\Gamma}_b^{-1}\approx40$\,s,
which is in good agreement with the independent measurements of the decoherence rate.
Although the magnetic responsivity deteriorates in the deamplification regime,
it is well suited for suppressing environmental \mbox{magnetic noise by at least one order of magnitude}.

\begin{figure*}[t]  
	\makeatletter
	\def\@captype{figure}
	\makeatother
	\includegraphics[scale=1.05]{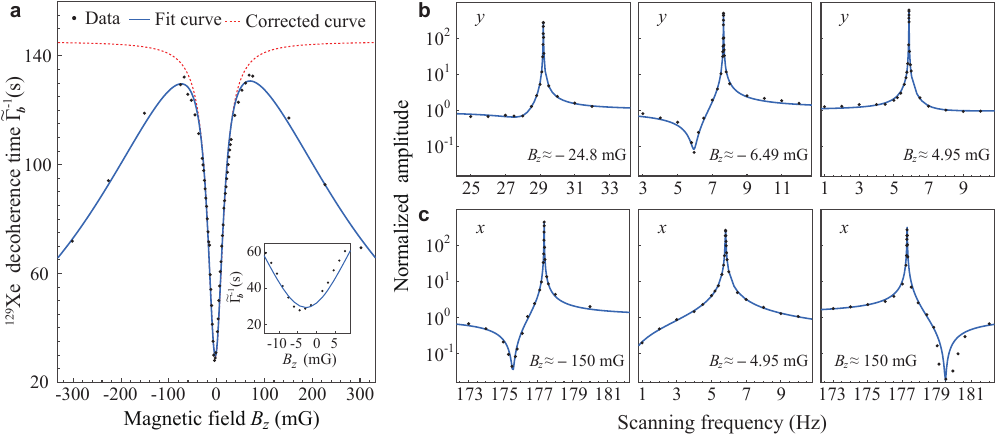}
	\caption{\textbf{Control of Fano resonance with an external magnetic field.} \textbf{a}, $\rm^{129}Xe$ decoherence time as a function of external magnetic field. The $\rm^{129}Xe$ decoherence time reaches a minimum $\widetilde{\Gamma}_{b}^{-1}\approx 28$\,s at $B_z \approx -3$\,mG and reaches a maximum $\widetilde{\Gamma}_{b}^{-1}\approx 130$\,s at $B_z \approx \pm 70$\,mG. After increasing external field, the decoherence time decreases due to magnetic field gradient. The solid line is the theoretical fit with our built noble-gas decoherence model, and the dashed line represents the $\rm^{129}Xe$ decoherence after numerically correcting the magnetic field gradient. \textbf{b} and \textbf{c}, Fano profiles for $x,y$-direction oscillating fields under different bias fields.}
	\label{Figure3}
\end{figure*}

Fano resonance can be controlled with an external magnetic field, enabling tuning the sensing performance.
Diagonalizing the matrix in Eq.\,(\ref{eq1}), we find the complex energy eigenstates of the 
the hybrid spin system:  $\widetilde{\omega}_{a,b}+i\widetilde{\Gamma}_{a,b}=\omega_0+i\chi\pm\sqrt{J^2+\Gamma^2}$, where $\omega_0=(\omega_a+\omega_b)/2$, $\chi=(\Gamma_a+\Gamma_b)/2$ and $\Gamma=\delta+i\beta$ with $\delta=(\omega_a-\omega_b)/2$, $\beta=(\Gamma_a-\Gamma_b)/2$ (Supplementary Section\,\uppercase\expandafter{\romannumeral1}).
We show that the joint action of magnetic Zeeman interaction and the alkali-metal-noble-gas bidirectional coupling dresses their Larmor frequencies and decoherence rates, in particular $\widetilde{\Gamma}_b^{-1}$ of $\rm^{129}Xe$ spins.
In experiment,
the decoherence time reaches a minimum $\widetilde{\Gamma}_{b}^{-1}\approx 28$\,s when the bias field is $B_z \approx -3$\,mG.
Such a minimum can be theoretically estimated by calculating the extreme value of $\widetilde{\Gamma}_{b}^{-1}$ and satisfies $\delta=0$, where the external field is $B_z =\lambda(\gamma_b M^a_z-\gamma_a M^b_z)/(\gamma_a-\gamma_b) \approx -\lambda M^b_z$.
In this case, the electron and nuclear spin precession frequencies are nearly matched, leading to the strong damping of $\rm^{129}Xe$ precession.
The $\rm^{129}Xe$ decoherence time increases rapidly when $B_z$ is increased.
When the bias field is $B_z\approx\pm70$\,mG, the decoherence time reaches a maximum $\widetilde{\Gamma}_{b}^{-1}\approx 130$\,s.
We found that with a further increase of $B_z$,
the $^{129}$Xe decoherence time decreases, in contradiction with the expectation [Fig.\,\ref{Figure3}(a)]. This is caused by the inhomogeneity of the applied magnetic field.
In order to verify this,
we built a comprehensive model of noble-gas spin decoherence caused by bidirectional coupling and magnetic field gradient (Supplementary Section\,\uppercase\expandafter{\romannumeral2}).
With this decoherence model, the data are well fit [solid curve in Fig.\,\ref{Figure3}(a)].
Based on our analysis, the magnetic field inhomogeneity is estimated to be about 0.04\%.
We numerically correct the magnetic gradient effect and obtain the corrected decoherence with the dashed curve in Fig.\,\ref{Figure3}(a).
The width of the dashed curve is proportional to the $\rm ^{87}Rb$ decoherence rate $\Gamma_a$ and
we notice that the decoherence rate could be recovered to that of uncoupled $\rm ^{129}Xe$ spins, i.e., $\widetilde{\Gamma}_{b}\approx\Gamma_b$ at the bias field larger than about 70\,mG (Supplementary Section\,\uppercase\expandafter{\romannumeral1}).

When increasing external magnetic field, a different regime emerges where Fano resonance of the oscillating field along $x$ becomes significant.
The Fano profile for the measured field along $y$ gradually becomes symmetric as $B_z$ increases [Fig.\,\ref{Figure3}(b)].
By contrast, using the same method described above,
we observe significantly asymmetric Fano profile [Fig.\,\ref{Figure3}(c)] when a $x$-direction oscillating is applied at large $B_z$, for example, $B_z\approx\pm 150$\,mG in our experiment.
The magnetic responsivity is enhanced over 200 at the amplification regime and suppressed over 10 at the deamplification regime.
It is worth noting that the Fano profile is reversed in Fig.\,\ref{Figure3}(c) due to the sign change of  the Fano parameter $q$ depending on the direction of $B_z$.

\begin{figure*}[t]  
	\makeatletter
 \centering
	\def\@captype{figure}
	\makeatother
	\includegraphics[scale=1.17]{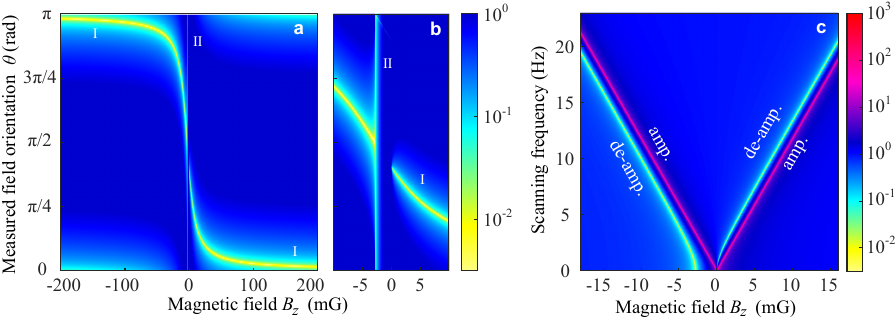}
	\caption{\textbf{Sensing with optimized amplification and deamplification.} \textbf{a} and \textbf{b}, Deamplification at various external magnetic fields and sensing directions. For each magnetic field $B_z$, we scan the direction of measured field from 0 to $\pi$ and record the minimum of Fano profile that is normalized comparing with the response far away from $^{129}$Xe Larmor frequency (in our experiment, we choose 300\,Hz response as a reference). There are two distinct curves, which present optimal sensing directions for realizing significant deamplification. The curve \uppercase\expandafter{\romannumeral1} corresponds to the Fano resonance near the $\rm ^{129}Xe$ Larmor frequency and is broken near the strong-damping point $B_z\approx -\lambda M^b_z$ without deamplification. The curve \uppercase\expandafter{\romannumeral2} is sharp at $B_z\approx -\lambda M^b_z$. \textbf{c}, Amplification and deamplification along the optimal sensing direction.}
	\label{Figure5}
\end{figure*}

We further investigate sensing with optimized operation parameters, including sensing direction and external magnetic field.
We numerically simulate the Fano profile of oscillating field with various directions described with the azimuth angle $\theta$ ranging from 0 to $\pi$ (rotating along $z$ in the $xy$ plane),
and then scan the external field frequency (Supplementary Section\,\uppercase\expandafter{\romannumeral5}).
As one can see, there are two distinct curves (\uppercase\expandafter{\romannumeral1}, \uppercase\expandafter{\romannumeral2}) in Fig.\,\ref{Figure5}(a),
where we focus on the minimum value for each Fano profile.
Our results show that the curve \uppercase\expandafter{\romannumeral1} is thicker than curve \uppercase\expandafter{\romannumeral2}, yielding broader direction range for implementing deamplification.
We find that the curve \uppercase\expandafter{\romannumeral1} can be approximated with $\theta\approx\cot^{-1}\gamma_a(B_z+\lambda M^a_z +\lambda M^b_z)/\Gamma_a$, which is the optimal direction for performing deamplification-assisted metrology.
In this case, the detection noise could be suppressed by a factor of more than 100. 
The curve \uppercase\expandafter{\romannumeral1} is disconnected at a small field range [Fig.\,\ref{Figure5}(b)] because the deamplifiction point is moved to negative frequencies that can not be observed.
The curve \uppercase\expandafter{\romannumeral2} is extremely sharp as a function of external magnetic field,
which is close to a specific value $B_z=-\lambda M^b_z-\lambda M^a_z$.
At this field, the deamplification works well for arbitrary measured-field directions from 0 to $\pi$ in the $xy$ plane.
We note that such $B_z$ is the operation field of self-compensated comagnetometers\,\cite{kornack2002dynamics,vasilakis2009limits,liu2022dynamics}.

Figure\,\ref{Figure5}(c) shows the performance of optimal sensing with optimized sensing direction.
Here we focus on the curve \uppercase\expandafter{\romannumeral1} that corresponds to the high-frequency part of the Fano profile.
The amplification and deamplification points are linearly dependent on the bias field when the bias field is far away from the self-compensating point [Fig.\,\ref{Figure5}(c)]. 
We further show that the frequency distance between such two points remain a constant  $\Delta \nu \approx \gamma_b\lambda M^b_z/(4\pi)$ (Supplementary Section\,\uppercase\expandafter{\romannumeral4}).
In such optimal conditions, the deamplification factor can be higher than 100,
which is about one order of magnitude better than when the field is along the non-optimal $x$ or $y$ directions.
By contrast, the amplification is independent of azimuth angle $\theta$.
Inspired by the above finding,
we could adjust the relative direction between the sensor and measured field to optimize the metrological performance before sensing.

The present amplification and deamplification provide two flexible methods to bypass the low-noise requirement for detection of small signals.
The type of detection noise is the key criterion to select amplification or deamplification.
On the one hand, when the dominant noise does not directly interact with noble-gas spins, for example, probe photon-shot noise and alkali-metal spin-projection noise\,\cite{budker2007optical,kominis2003subfemtotesla,bao2020spin}, such detection noise is not amplified with noble gas.
In this case, the amplification provides an efficient way to enhance the sensitivity by the amplification factor compared with the sole alkali-metal magnetometer.
In our experiment,
the magnetic sensitivity is enhanced to about 3.5\,fT/Hz$^{1/2}$ assisted with the amplification of more than 500,
which is about 54\,dB below the dominant photon-shot noise in $^{87}$Rb magnetometer.
Additionally, the achieved sensitivity is equivalent to the energy resolution of  $2.7\times 10^{-23}$\,eV/Hz$^{1/2}$.
Our method does not use squeezed light, relaxing the implementation complexity.
On the other hand, when the detection noise simultaneously interact with noble gas and alkali metal with the scale of their magnetic moment (for example, magnetic noise), it is better to choose the deamplification operation.
The deamplification-assisted sensitivity is equal to the alkali-metal sensitivity.
Such a metrology sensitivity is about 1\,pT/Hz$^{1/2}$ with $\rm ^{87}Rb$-$^{129}$Xe system in our experiment (Supplementary Section\,\uppercase\expandafter{\romannumeral6}).
However, the use of K-$\rm ^{3}He$ system could improve such a sensitivity, where K magnetometer has already demonstrated fT/Hz$^{1/2}$-level sensitivity that is typically limited by magnetic noise from magnetic shields\,\cite{kominis2003subfemtotesla,budker2007optical}.
The deamplification offers the capability to beat the magnetic noise limit.
In actual, previous self-compensated comagnetometer is one type of deamplification-assisted sensors and has demonstrated 0.75\,fT/Hz$^{1/2}$ at about 0.1\,Hz\,\cite{vasilakis2009limits}.
Our work opens new opportunities to extend the sub-femtotesla and even higher sensitivity to higher frequency range, for example, 1-100\,Hz.

We would like to emphasize the difference between this work and body of existing work.
In Ref.\,\cite{jiang2021search},
we reported on using noble gas as a spin amplifier to enhance magnetic sensing.
This work further gains the deep understanding of spin amplification with Fano resonance.
Moreover, beneficial from the new connection between Fano resonance and coupled atomic gases,
we find unexplored deamplification phenomena in a broad frequency range.
Although there have been works on self-compensated comagnetometers\,\cite{kornack2002dynamics,vasilakis2009limits,liu2022dynamics},
they only show the deamplification of magnetic noise in near-zero frequency range.
By contrast,
our work builds a unified framework of self-compensating mechanism based on Fano resonance that can explain previous near-zero-frequency self-compensating mechanism and moreover extends its operation to high-frequency range for the first time, for example, above 100 Hz.
Our advances would open new opportunities for enhanced quantum metrology in a wide range of precision measurements\,\cite{balser1960observations,jiang2021search,afach2021search,arvanitaki2014resonantly,bloch2022new}.

Our work may advance enhanced quantum sensing by bridging non-Hermitian physics and coupled atomic gases.
According to Eq.\,(\ref{eq1}),
the dynamics of alkali-metal and noble-gas excitations is governed by a non-Hermitian Hamiltonian,
where there is an exceptional point （EP）
corresponding to $\delta=0$ and \mbox{$J/\beta=1$} (Supplementary Section\,\uppercase\expandafter{\romannumeral1}).
In the vicinity of the EP where a small perturbation from external field causes non-zero $\delta$,
the frequency splitting between alkali-metal and noble gas resonance lines becomes $\Delta\widetilde{\omega}_{a,b}=2\beta^{1/2}|\delta|^{1/2}$.
This sublinear response leads to a sizable enhanced sensing compared with the linear response relying on Hermitian degeneracies,
with an amplification factor of $(\beta/\delta)^{1/2}\gg 1$.
Yet, the key realization of $J/\beta=1$ is still out of reach in $\rm ^{87}Rb$-$\rm ^{129}Xe$ system,
where $J/\beta\approx10^{-3}$ in our experiment.
By contrast,
the proposed EP-enhanced sensing should be realized in $\rm K$-$\rm ^{3}He$ coupled gases,
where $J/\beta$ has been recently demonstrated from 1-20\,\cite{shaham2022strong,kornack2002dynamics}.

Our work suggests a novel quantum-sensing technique to search for hypothetical particles beyond the standard model\,\cite{preskill1983cosmology,abbott1983cosmological,dine1983not,dobrescu2006spin}
such as ultralight axions or dark photons that are also well-motivated dark matter candidates.
Such particles couple with standard-model particles (for example, to nuclei interacting with their spins) or act as force mediators to induce exotic interactions between two standard-model particles.
Here the hypothetical particles are assumed to couple with only one of the alkali-metal and noble-gas spins.
As a consequence, hypothetical particles produce an oscillating pseudo-magnetic field on spins.
The sensors presented in this work are, in effect, transducers converting the pseudo-magnetic field into an effective magnetic field that is measured.
For example, the pseudo-magnetic field on noble-gas spins can be effectively magnified by almost three orders of magnitude.
On the other hand, deamplification-assisted sensor can extend the previous applications of self-compensated comagnetometers to, for example, exotic spin-dependent interactions with high-modulation frequencies.
With our current experimental parameters,
the search sensitivity of axions and dark photons is well beyond the most stringent supernova constraints by about two orders of magnitude\,\cite{raffelt2008astrophysical}.
Our method can also be used to measure weak magnetization of samples, for example, ancient rocks for archaeometry, where rocks are rotated within the amplification frequency range. 
In contrast to superconducting quantum interference devices that are dominant tools in archaeometry,
our method does not require cryogenic cooling.

In summary, we report the first demonstration of Fano resonance in coupled alkali-metal-noble-gas spin system and implement enhanced metrology with relaxed requirements on the detection noise (as the signal is selectively amplified).
Our work simultaneously provides amplification and deamplification as two critical resources for enhanced sensing, which does not require quantum entanglement.
Our sensing concept is generic and can be extended to other well-established Fano resonance in, for example, metamaterials, plasmonic-atomic system, and nuclear scattering\,\cite{limonov2017fano,miroshnichenko2010fano,luk2010fano}, enabling exciting applications in sensing electromagnetic fields from microwave, optics to X ray.

\bibliographystyle{naturemag}
\bibliography{mainrefs}

\noindent
\textbf{Code availability}.
The code that supports the plots in this paper is available from the corresponding author upon reasonable request.

~\

\noindent
\textbf{Acknowledgements}. This work was supported by the Innovation Program for Quantum Science and Technology (Grant No. 2021ZD0303205),
National Natural Science Foundation of China (grants nos. 11661161018, 11927811, 12004371, 12150014, 12205296, 12274395), Youth Innovation Promotion Association (Grant No. 2023474).
This work was also supported by the Cluster of Excellence ``Precision Physics, Fundamental Interactions, and Structure of Matter'' (PRISMA+ EXC 2118/1) funded by the German Research Foundation (DFG) within the German Excellence Strategy (Project ID 39083149).

~\

\noindent
\textbf{Author contributions}.
M.J., Y.S.Q. designed experimental protocols, analyzed the data and wrote the manuscript.
Y.H.W., Y.H. analyzed the data and edited the manuscript.
X.H.P., D.B. proposed the experimental concept, devised the experimental protocols, and edited the manuscript.
All authors contributed with discussions and checking the manuscript.

~\

\noindent
\textbf{Competing interests}.
The authors declare no competing interests.

\end{document}